\documentclass[preprintnumbers,amsmath,11pt,amssymb,floatfix,superscriptaddress,nofootinbib]{article}

\def\mytitle#1{\setcounter{equation}{0}
\setcounter{footnote}{0}
\begin{flushleft}\Large\textbf{#1}\end{flushleft}
\vspace{0.25cm}}
\def\myname#1{\leftline{{\large #1}}\vspace{-0.13cm}}
\def\myplace#1#2{\small\begin{flushleft}\textit{#1}\\
\texttt{#2}\end{flushleft}}

\def\myclassification#1{\small\noindent
Keywords :
       #1\vspace{0.5cm}}
\usepackage{graphicx}
\begin{document}
\mytitle{Thermoynamics of $d-$Dimensional Charged AdS (Anti-de Sitter) Black Holes: Hamiltonian Approach and Clapeyron Equation}

\myname{$Amritendu~ Haldar^{*}$\footnote{amritendu.h@gmail.com} and $Ritabrata~
Biswas^{\dag}$\footnote{biswas.ritabrata@gmail.com}}
\myplace{*Department of Physics, Sripat Singh College, Jiaganj, Murshidabad $-$ 742123, India.\\$\dag$Department of Mathematics, The University of Burdwan, Golapbag Academic Complex, City : Burdwan  $-$ 713104, District : Purba Burdwan, State : West Bengal, India.} {}
 
\begin{abstract}
The study of thermodynamics in the view of the Hamiltonian approach is a newest tool to analyze the  thermodynamic properties of the black holes. In this letter, we investigate the thermodynamics of $d$-dimensional ($d>3$) asymptotically AntideSitter black holes. A thermodynamic representation based on symplectic geometry is introduced in this letter. We extend the thermodynamics of $d-$dimensional charged AntideSitter black holes in the views of a Hamiltonian approach. Firstly, we study the thermodynamics in reduced phase space and correlate with the Schwarzschild solution. Then we enhance it in the extended phase space. In an extended phase space the thermodynamic equations of state are stated as constraints. We apply the canonical transformation to analyze the thermodynamics  of said type of black holes. We plot $P$-$v$ diagrams for different dimensions $d$ taking the temperatures $T<T_c, T=T_c $ and $T>T_c$ and analyze the natures of the graphs and the dependences on $d$. In theses diagrams, we point out the regions of coexistence. We also examine the phase transition by applying ``Maxwell's equal area law" of the said black holes. Here we find the regions of coexistence of two phases which are also depicted graphically. Finally, we derive the ``Clapeyron equation" and investigate the latent heat of isothermal phase transition.
\end{abstract}

\myclassification{Maxwell's equal area law; Clapeyron equation and latent heat.}


\section{Introduction :}
Hawking put forward his idea that the radiation can be found to come out of black holes (BHs hereafter) via quantum tunneling \cite{Bardeen 1973, Hawkings 1974}. This idea provides a real connection with quantum mechanics to gravity. The main consideration of BHs is that thay have a certain geometrical boundary the name of which is coined as event hotizon. Actually the thermodynamics of BHs is the construction of analogus thermodynamic parameters which are obtained on the event horizon. In the references \cite{Page 1983} the authors suggest that the phase transition called Hawking-Page phase transition. These kind of transition be explained as the confinement/deconfinement of phase transition of gauge field in the AntideSitter (AdS hereafter)/ conformal field theory (CFT) correspondence \cite{Witten 1998, Maldacena 1998, Gubser 1998}. In broader sense, the AdS/ CFT correspondence is associated with gravity physics in an asymptotically AdS geometry with field theory in the boundary surface of the AdS space. The authors of the referenece \cite{Gubser 1998} suggest extended gauge/ gravity dualities in context with string theory.

In recent years the thermodynamic aspects of AdS space-times are considered in several references like \cite{Brown 1994, Louko 1996, Hemming 2007, Rajeev 2008, Miranda 2009, Hubeny 2010, Banerjee 2012, Menoufi 2013, Myung 2014, Lemos 2015, biswas1, biswas2, biswas3, biswas4, biswas5, biswas6, Chowdhury1, Haldar 2018, Haldara 2018}. The thermodynamics of the BHs with cosmological constant $\Lambda$ as a thermodynamical variable has been studied in the references \cite{Teitelboim 1985, Sekiwa 2006, Urano 2009}. The physical interpretation of the conjugate variable associated with $\Lambda$ is a point of interest till now. Though in the references \cite{Dolan 2011, Dolan 2012} the authors suggest that this conjugate variable could be thermodynamic volume and due to this proposal $\Lambda$ would be interpreted as a pressure and the mass of the BHs would be identified with the total gravitational enthalphy and not with the internal energy of the system in the extended phase space.

After development of string theory or M-theory, the study of the higher dimensional BHs comes in interest.  
This theory plays a very important role to build the quantum theory of gravity. The study of dynamics of BHs in higher dimensions gives the interesting result even in studying the compactified mechanisms. In \cite{Strominger 1996, Breckenridge 1997} the authors has established the statistical- mechanical calculations of the Bekenstein-Hawking entropy for a class of supersymmetric BHs in 5-dimensions. This is one of the remarkable results in string theory. Brane world scenarios also motivate to study the BHs in higher dimensions. This is a new fundamental scale of quantum gravity. As the dimensions of space-time affect the thermodynamic properties of BHs \cite{Davies 1989, Emparan 2000, Myung 2007, Chakraborty
2015}, the analyze of BHs in higher dimensions is very much significant.

In this letter, we consider the thermodynamic aspects of AdS space-time. Here we choose the $d$-dimensional($d>3$) charged AdS BHs whose metric is the spherically symmetric solution of  Einstein field equations (EFE) with a negative cosmological constant \cite{Fan-2016}. We introduce the cosmological constant as thermodynamic variable as other existing works viz., \cite{Teitelboim 1985, Sekiwa 2006, Urano 2009}. Only deference with these articlas is that they have taken $\Lambda$ to be the function of coordinates in phase space through a new equation of state. Surprisingly this new equation of state exhibits the generalized thermodynamic volume. In this regard we study the thermodynamics of $d-$dimensional charged AdS BHs in view of the Hamiltonian approach as done by the authors of references \cite{Baldiotti 2016, Baldiotti 2017}. A sympletic geometry will be used to describe the thermodynamics and we realize that the thermodynamic equation of states are constraints on phase space. This new treatment which is introduced here is consistent because it resembles with the new thermodynamic potential and respects the laws of BHs thermodynamics. We may construct a new equation of state which completely characterizes the thermodynamics of charged$d-$ dimensional charged AdS BHs only assuming homogeneity of the thermodynamic variables. 

Using ``Maxwell's equal area law", we analyze  the phase transition of the said BHs. Here we notice that there is the regions of coexistence  of two phases. We will also depict the coexistence region of two phases by plotting the $P$-$v$ diagrams for different dimensions $d$, for the temperatures $T<T_c, T=T_c $ and $T>T_c$ respectively. Finally, we derive the ``Clapeyron equation" and investigate the latent heat of isothermal phase transition by studying the phase transition process.

The letter is organized as follows. In section $2$ we establish the thermodynamic results of the $d-$dimensional charged AdS (Anti-de Sitter) BHs in brief. We also study the thermodynamics of that   BHs in reduced as well as extended phase space in the view of a Hamiltonian approach in the same section. In section $3$ we extend `Maxwell's' equal area law for this BHs. We compute the `Clapeyron equation' and investigate the latent heat of isothermal phase transition in section $4$. Finally we briefly discuss about the out come of this letter and conclude this letter in section $5$. Through out this letter we use the Planck units, i.e.,  $G =c = \hbar = \kappa_B = 1$ and signature $(-++.....+)$.    

\section{The Brief Review of Thermodynamics of $d-$dimensional Charged AdS Black Holes:}

The $d-$dimensional ($d>3$) charged AdS BHs metric can be obtained as the spherically symmetric solution of d-dimensional Einstein field equation (EFE) of the form:  $ G_{\mu\nu}+ \Lambda g_{\mu\nu}= 8\pi T_{\mu\nu} $ \cite{Gunasekaran 2012a}:
\begin{equation}\label{ah8.equn1}
ds^2=-g_{tt}(r)dt^2+g_{rr}dr^2+ r^2\bigg\{(d\theta_1)^2+sin^2\theta_1(d\theta_1)^2+.....+sin^2\theta_1 ...... sin^2\theta_{d-2}(d\theta_{d-2})^2\bigg\},
\end{equation} 
where 
\begin{equation}\label{ah8.equn2}
-g_{tt}(r)=g_{rr}(r)^{-1}=1-\frac{\tilde{M}}{r^{d-3}}+\frac{\tilde{Q}^2}{r^{2(d-3)}}-\tilde{\Lambda}r^2 
\end{equation}
and
$(d\theta_1)^2+sin^2\theta_1(d\theta_1)^2+......+sin^2\theta_1 ...... sin^2\theta_{d-2}(d\theta_{d-2})^2$ represents the canonical volume associated with the induced metric (\ref{ah8.equn1}) and
the constants $\tilde{M}$, $\tilde{Q}$ and $\tilde{\Lambda}$ in the $equation$ (\ref{ah8.equn2}) are expressed in terms of the mass $M$, the electrical charge and the cosmological constant $\Lambda$ respecively \cite{Baldiotti 2017} as:
\begin{equation}\label{ah8.equn3}
\tilde{M}=\frac{16\pi M}{(d-2)B_{d-2}},~~~ B_{d-2}=\frac{2\pi^\frac{d-1}{2}}{\Gamma\left(\frac{d-1}{2}\right)},~~~~~\tilde{Q}=\frac{8\pi Q}{B_{d-2}\sqrt{2(d-3)(d-2)}} ~~~~~and ~~~~\tilde{\Lambda}=\frac{2\Lambda}{(d-2)(d-1)}~.
\end{equation}
On the event horizon $r=r_h$, the mass parameter $\tilde{M}$ can be expressed as:

\begin{equation}\label{ah8.equn4}
{\tilde{M}}=\frac{\tilde{Q}^2}{r_h^{d-3}} +\left(1-\tilde{\Lambda} r_h^2\right)r_h^{d-3}.
\end{equation}
Assuming that the event horizon is a Killing horizon, the surface gravity may be defined as the magnitude of the norm of horizon generating Killing field $\chi^a=\zeta^a+\Omega\psi^a$, evaluated at the horizon and is expressed by using equations (\ref{ah8.equn2}) and (\ref{ah8.equn4}) as:

\begin{equation}\label{ah8.equn5}
\kappa=lim_{r\longrightarrow r_h}\frac{1}{2}\frac{d}{dr}\left[\sqrt{-{g_{tt}(r)\times g_{rr}(r)^{-1}}}\right]=\frac{1}{2}\left[\frac{d-3}{r_h}-\tilde{\Lambda}(d-1)r_h-\frac{\tilde{Q}^2 (d-3)}{r_h^{2d-5}}\right].  
\end{equation}
The Killing horizon area $A$ may be written as a function of event horizon's radius as: 
\begin{equation}\label{ah8.equn6}
A=r_h^{d-2}B_{d-2}.
\end{equation} 
Stationary BH solutions of EFE relate the first law of thermodynamics as:
\begin{equation}\label{ah8.equn7}
dM=T dS + \Omega dJ + \phi dQ
\end{equation}
and the Smarr formula for $d-$dimensional charged AdS BHs reads
\begin{equation}\label{ah8.equn8}
M=\frac{d-2}{d-3} TS  + \phi dQ - \frac{2}{d-3} PV.
\end{equation}
The entropy $ S $ of BHs is related to the area $ A $ at event horizon as:
\begin{equation}\label{ah8.equn9}
S=\frac{A}{4}=\frac{r_h^{d-2} B_{d-2}}{4} 
\end{equation}
and the electric potential for this kind of BHs is given as:
\begin{equation}\label{ah8.equn10}
\phi = \frac{\tilde{Q}^2}{r_h^{d-3}} \sqrt{\frac{d-2}{2(d-3)}}.  
\end{equation}
The classical BH thermodynamics \cite{Bekenstein-1974, Hawking-1975, Hawking-1976} shows that the mass $ M $ and surface gravity $ \kappa $ of the BHs related to internal energy $ U $ and  temperature $ T $ as:
\begin{equation}\label{ah8.equn11}
 M=U ~~~~~~~~   and ~~~~~~~~  T=\frac{\kappa}{2\pi}. 
\end{equation}
More generally the mass $ M $ of the BHs is equal to the total gravitational enthalpy $ H $ \cite{Fan-2016} and is  expressed as:
\begin{equation}\label{ah8.equn12}
M=H=U+PV ,
\end{equation}
where $ P $ is the thermodynamic pressure and $ V $ is the naive geometric volume and they are defined as:
\begin{equation}\label{ah8.equn13}
P=-\frac{\tilde{\Lambda}(d-2)(d-1)}{16\pi}
~~~~~~~~~ and ~~~~~~~~ V=\frac{1}{d-1}\left(\frac{4S}{B_{d-2}}\right)^\frac{d-1}{d-2}B_{d-2}.
\end{equation} 
Again
\begin{equation}\label{ah8.equn14}
dH=dM=TdS+VdP ,
\end{equation}
which gives,
\begin{equation}\label{ah8.equn15}
V=\left(\frac{\partial H}{\partial P}\right)_S =\left(\frac{\partial M}{\partial P}\right)_S
~~~~~~~~~~~ and~~~~~~~~~~~~T=\left(\frac{\partial H}{\partial S}\right)_P =\left(\frac{\partial M}{\partial S}\right)_P. 
\end{equation}
This volume $ V $ is known as thermodynamic volume. Generally thermodynamic volume is not equal to the naive geometric volume of the BHs.\\
Using equations (\ref{ah8.equn3}), (\ref{ah8.equn4}) and (\ref{ah8.equn6}) one can obtain the mass $M$ as:

\begin{equation}\label{ah8.equn16}
M=\frac{(d-2)B_{d-2}}{16\pi}\left(\frac{A}{B_{d-2}}\right)^{\frac{d-3}{d-2}}\left[\frac{\tilde{Q^2}}{\left(\frac{A}{B_{d-2}}\right)^{\frac{2(d-3)}{d-2}}}+1-\tilde{\Lambda}\left(\frac{A}{B_{d-2}}\right)^{\frac{2}{d-2}}\right].
\end{equation}
Moreover, using equations (\ref{ah8.equn5}) and (\ref{ah8.equn6}) one can construct the surface gravity $\kappa$ as:
\begin{equation}\label{ah8.equn17}
\kappa=\frac{1}{2} \left(\frac{A}{B_{d-2}}\right)^{\frac{1}{d-2}} \left[ \frac{d-3}{\left(\frac{A}{B_{d-2}}\right)^{\frac{2}{d-2}}} - \tilde{\Lambda} (d-1) -\frac{\tilde{Q}^2 (d-3)}{\left(\frac{A}{B_{d-2}}\right)^2}\right].
\end{equation}
From equations (\ref{ah8.equn16}) and (\ref{ah8.equn17}) one can obtain the following relation,
\begin{equation}\label{ah8.equn18}
8\pi M\left(\frac{d-1}{d-2}\right)-\kappa A=A\left(\frac{B_{d-2}}{A}\right)^{\frac{1}{d-2}}+ \frac{(d-2) (d-1)B_{d-2}\tilde{Q}^2}{\left(\frac{A}{B_{d-2}}\right)^{\frac{d-3}{d-2}}}.
\end{equation}
It is shown that in equation (\ref{ah8.equn18}) the cosmological constant $\Lambda$ does not appear explicitly. Hence from equation (\ref{ah8.equn18}) we have the equation of state given as:
\begin{equation}\label{ah8.equn19}
U\left(\frac{d-1}{d-2}\right)-T S=\frac{1}{2\pi}\left(B_{d-2}\frac{S^{d-3}}{4}\right)^{\frac{1}{d-2}}-\frac{(d-2) (d-1)B_{d-2}\tilde{Q}^2}{8\pi \left(\frac{4S}{B_{d-2}}\right)^{\frac{d-3}{d-2}}}.
\end{equation}
Equation (\ref{ah8.equn12}) gives the internal energy as:

\begin{equation}\label{ah8.equn20}
U=\frac{(d-2)B_{d-2}}{16\pi}\left(\frac{4S}{B_{d-2}}\right)^{\frac{d-3}{d-2}}\left[1+\frac{\tilde{Q^2}}{\left(\frac{4S}{B_{d-2}}\right)^{\frac{2(d-3)}{d-2}}}\right].
\end{equation}
Equation (\ref{ah8.equn20}) is not compatible with the thermodynamic relation given by 
\begin{equation}\label{ah8.equn21}
T=\frac{\kappa}{2\pi}\neq\frac{\partial U}{\partial S},
\end{equation}
This implies that the thermodynamics defined by the relations (\ref{ah8.equn9}), (\ref{ah8.equn11}) and (\ref{ah8.equn13}) are not consistent. Since in this problem $V$ does not depend upon $P$, this problem is a consequence of the singularity of the Legendre transformation of the pair (P, V) \cite{Dolan 2011}.

Using the equations (\ref{ah8.equn11}), (\ref{ah8.equn13}) and (\ref{ah8.equn17}), one can obtain the equation of state for a $d-$dimensional charged AdS BH as:
\begin{equation}\label{ah8.equn22}
P= \frac{(d-2)T}{4 \left(\frac{A}{B_{d-2}}\right)^{\frac{1}{d-2}}}-\frac{(d-3)(d-2)}{16\pi \left(\frac{A}{B_{d-2}}\right)^{\frac{2}{d-2}}}+ \frac{\tilde{Q}^2 (d-3) (d-2)}{16\pi \left(\frac{A}{B_{d-2}}\right)^2}.
\end{equation}
With the specific volume $\it {v}=\frac{4 \left(\frac{A}{B_{d-2}}\right)^{\frac{1}{d-2}} l_p^{d-2}}{d-2} $, where $l_p= \sqrt{\frac{\hbar G}{c^3}}$ \cite{Gunasekaran 2012b}, we can compute the equation of state for this BH as:
\begin{equation}\label{ah8.equn23}
P= \frac{T}{v}-\frac{C}{v^2}+ \frac{D \tilde{Q}^2}{v^{2(d-2)}}~,
\end{equation}
where $C= \frac{d-3}{\pi (d-2)}$ and $D=\frac{2^{4(d-3)} (d-3)}{\pi (d-2)^{2d-5}}$.

\begin{figure}[h!]
\begin{center}
~~~~~~~~~~~~~~~~~~Fig.-1a~~~~~~~~~~~~~~~~~~~~~Fig.-1b~~~~~~~~~~~~~~~~~~~~~~~~~~~~~~Fig.-1c~~~~~~~~~~~\\
\includegraphics[scale=.5]{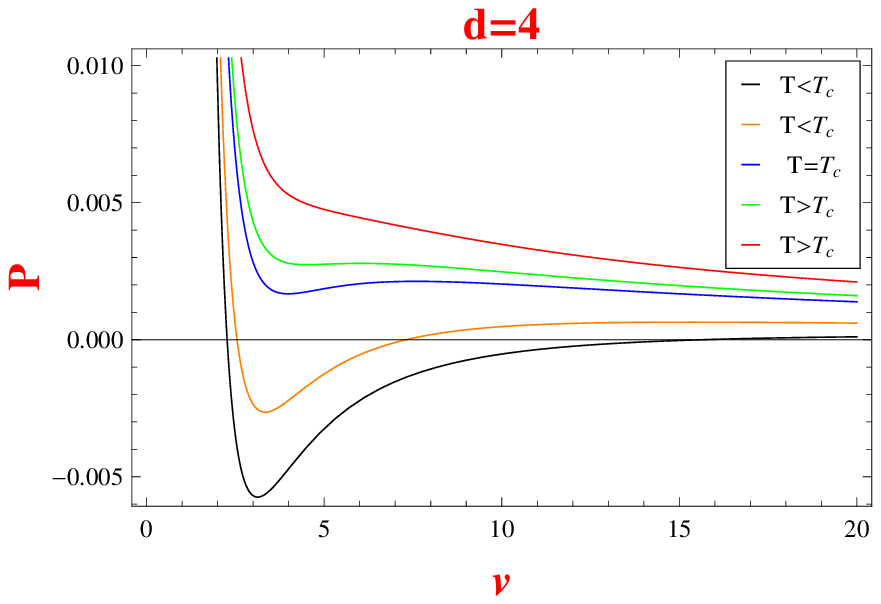}
\includegraphics[scale=.5]{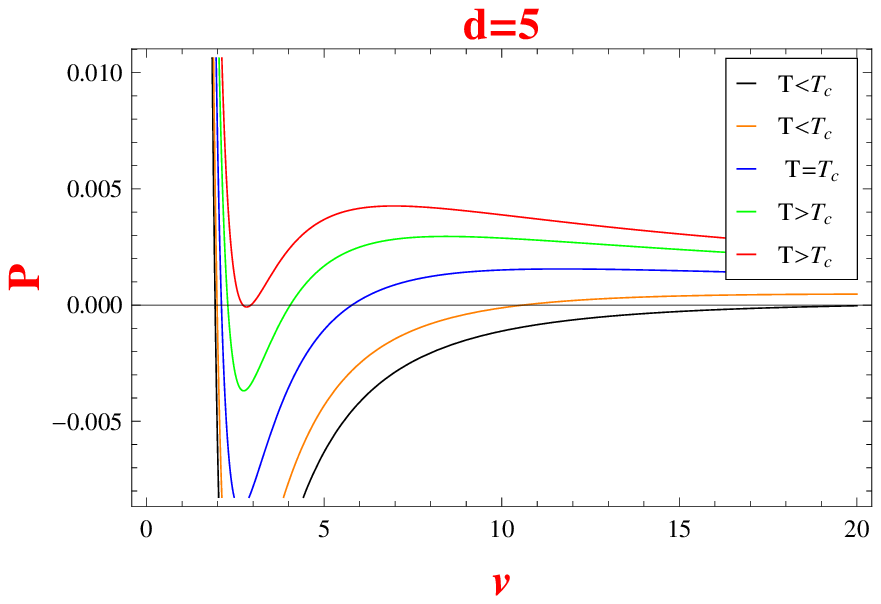}
\includegraphics[scale=.5]{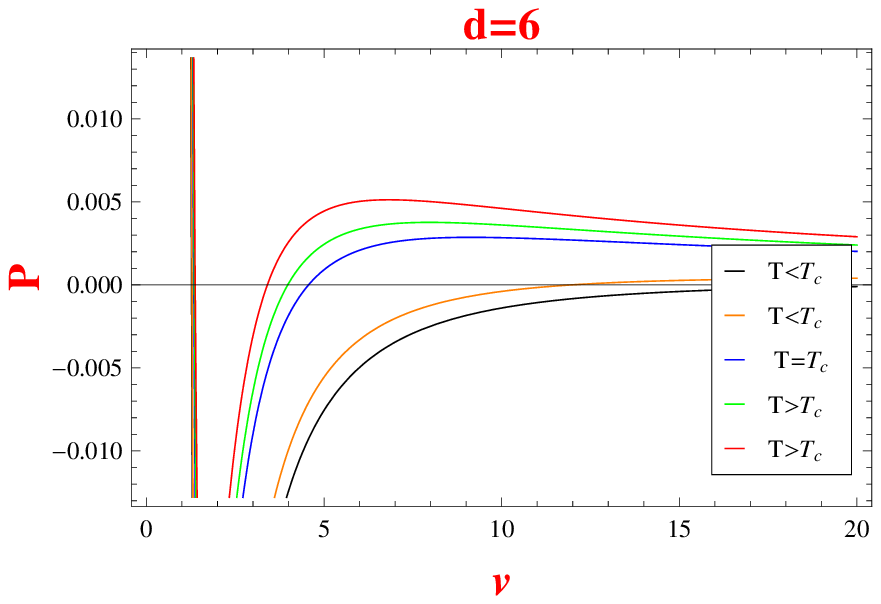}
~~~~~~~~~~~~~~~~~~Fig.-1d~~~~~~~~~~~~~~~~~~~~~Fig.-1e~~~~~~~~~~~~~~~~~~~~~~~~~~~~~~Fig.-1f~~~~~~~~~~~\\
\includegraphics[scale=.5]{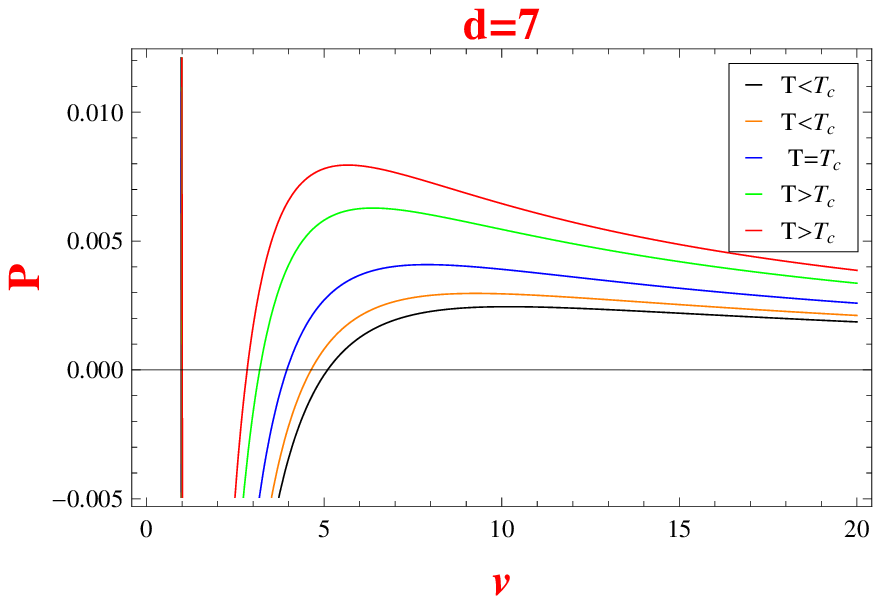}
\includegraphics[scale=.5]{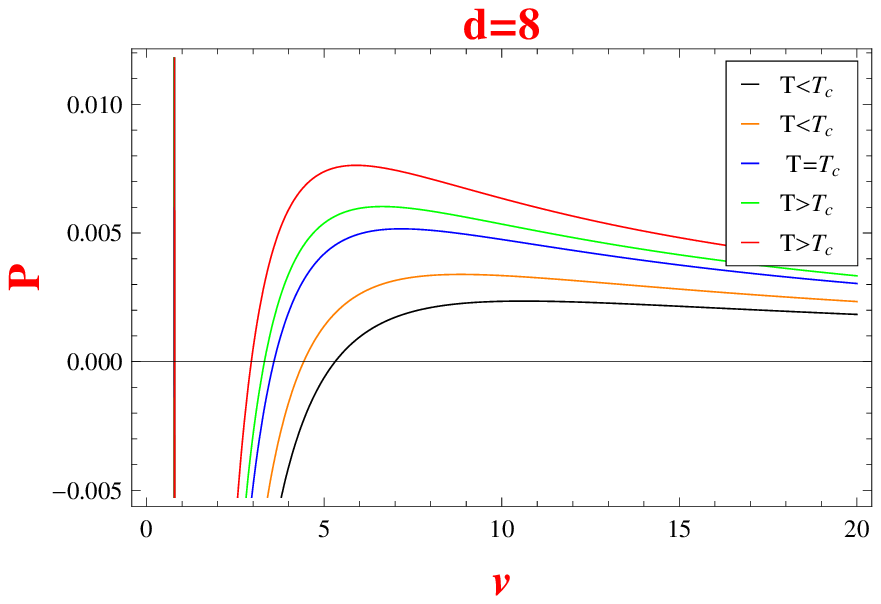}
\includegraphics[scale=.5]{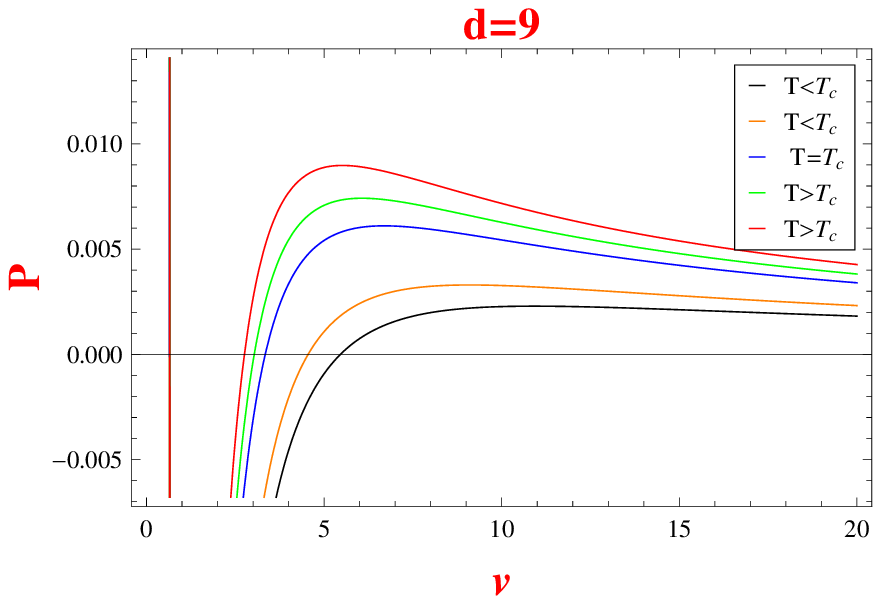}
~~~~~~~~~~~~~~~~~~~~~~~~Fig.-1g~~~~~~~~~~~\\
\includegraphics[scale=.5]{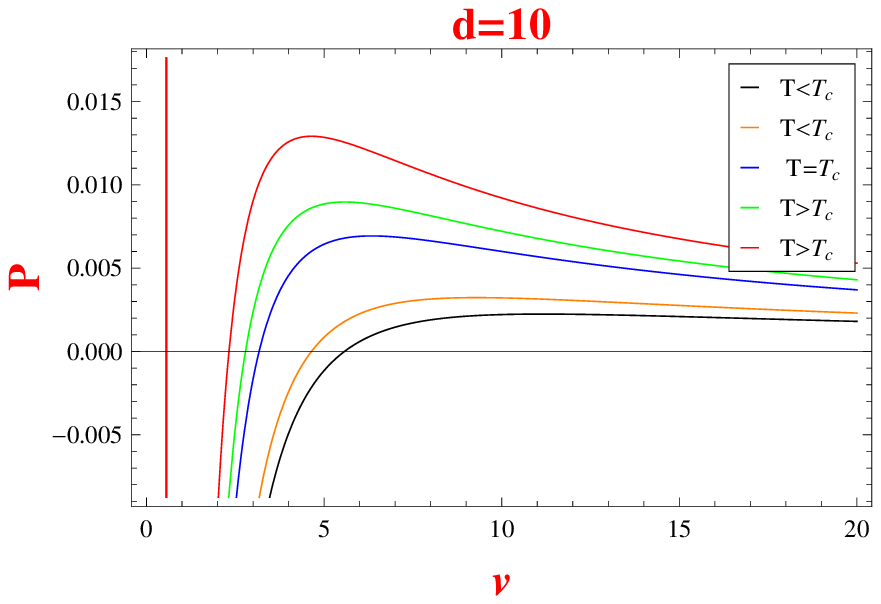}

Fig.-1a-g represent the variations of $ P $ with respect to specific volume $\it {v} $ for different dimensions $d$ for $T<T_c, T=T_c$ and $T>T_c$.
\end{center} 
\end{figure}

We have plotted the variations of $ P $ with respect to specific volume $\it {v} $ for different dimensions $d$ in Fig.-1a-g for equation (\ref{ah8.equn23}) for different dimensions $d$ for $T<T_c, T=T_c$ and $T>T_c$ respectively. Here we have seen that $P$ is decreased as $\it {v} $ is increased and reaches to a local minimum  for $T<T_c$. This means that there is a small-large BH phase transition in the system.  It is clearly shown in the curves that there are thermodynamic unstable regimes for $\frac{\partial P}{\partial v}> 0$ for $T<T_c$. For higher dimensions the thermodynamic unstable regimes also higher, i.e., for $T<T_c, T=T_c$ and for some extent of $T>T_c$ .

We also depict the $P$-$T$ diagrams in Fig.-2. Here we do not restrict the curves in critical points $(P_c~ and ~T_c)$. Here we notice that due to increase of temperature, pressure also increases but this increment is not linear as expected. It is also found that there are negative temperatures for $T<T_c$ which implies that there are thermodynamic unstable regimes for $T<T_c$. Again for higher dimensions this regimes also higher that also shown in Fig.-1a-g.
\begin{figure}[h!]
\begin{center}
~~~~~~~~~~~~~~~~~~Fig.-2~~~~~~~~~~~\\
\includegraphics[scale=.7]{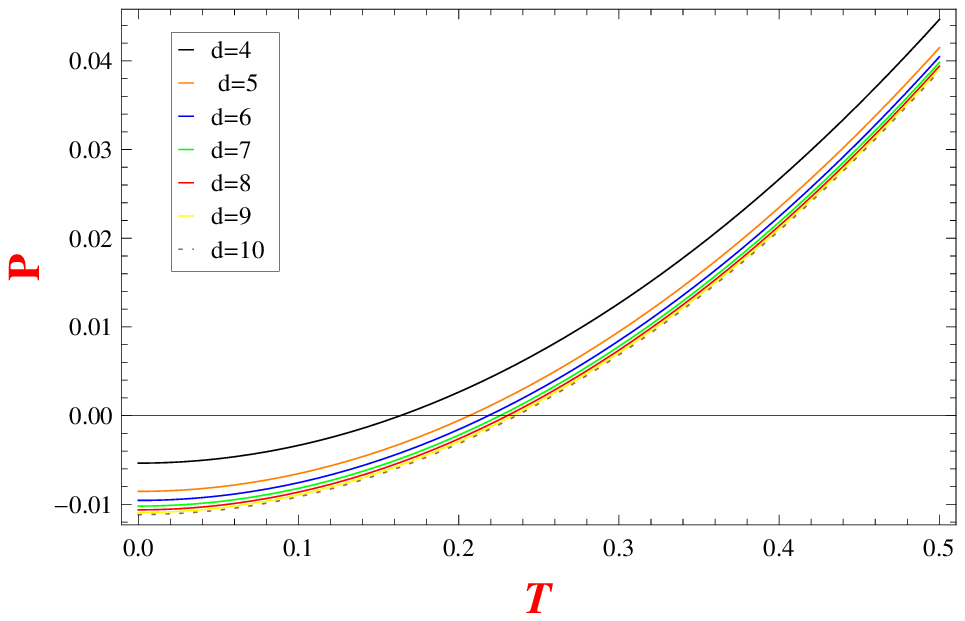}\\

Fig.-2 represents the variations of $ P $ with respect to the temperature $T $ for different dimensions $d$.
\end{center} 
\end{figure}
\subsection{Hamiltonian Approach to Thermodynamics of $d-$dimensional Charged AdS Black Holes:}
 Symplectic structure which involves thermodynamic variables are studied in the references \cite{Peterson 1979, Gambar 1994}. For some chosen phase space if the thermodynamic variables can be treated as the coordinates and momentum, the concerned thermodynamic variables' integrability conditions take place of the analogous Poission brackets and henceforth the duality between the mechanics and thermodynamics can be established. However, as it is not clear how to translate Hamiltonian trajectories in phase-space the analogy can not be taken too far. Translating Hamiltonian trajectories requires a clear idea of the Hamiltonian equations of motion $\dot{q}=\left\{q, H \right\}$, $\dot{p}=\left\{p, H \right\}$. In thermodynamics, the integrability conditions depend on the concerned potential chosen and they are related by Legendre transformations. Hamiltonian system is composed of the triple $(M, \omega, X_{H})$, $M$ is a smooth manifold, $\omega$ is the canonical symplectic form on $M$ and $X_{H}$ the Hamiltonian vector field.

All the equations of state of a thermodynamic system as constraints on phase space can be realised in the Hamiltonian approach to thermodynamics. On the constraints surface $(Hamiltonian ~(H)=0)$ with the relation $d\tau=dt,$ the tautological form in extended phase space degenerates to Poincar\'{e}-Cartan form in another phase space with reduced Hamiltonian, $h=h(q, p, \tau)$ as: $ p_i dq^i +\left.\zeta d \tau\right|_{H=0}=p_i dq^i -h dt,$ where $(\zeta,\tau)$ is a canonical pair by which one can extend the phase space \cite{Baldiotti 2016}. In this condition, one can simplify a Hamiltonian  system with Hamiltonian function $H=\zeta+h(q, p, \tau)$ to the system with $h$ in the reduced phase space $(q, p).$ The differential of  thermodynamic variables is given by the tautological form as: $ p dq +\zeta d \tau\equiv dU$ on the constraints surface. Thus in this way, all the thermodynamic variables are to be related by the canonical transformation and gives equivalent representations.

Here we introduce the canonical transformations as:        
\begin{equation}\label{ah8.equn24}
q= \frac{A}{B_{d-2}}= \frac{4S}{B_{d-2}}~~~~and~~~~p= \pi T= \frac{\kappa}{2} ,
\end{equation}
The equations (\ref{ah8.equn16}) and (\ref{ah8.equn17}) may be written in terms of the canonical transformations (\ref{ah8.equn24}) as:
\begin{equation}\label{ah8.equn25}
M=\frac{(d-2)B_{d-2}}{16\pi}q^{\frac{d-3}{d-2}}\left[\frac{\tilde{Q^2}}{q^{\frac{2(d-3)}{d-2}}}+1-\left\{\frac{2\Lambda}{(d-2)(d-1)}\right\}q^{\frac{2}{d-2}}\right].
\end{equation}
and 

\begin{equation}\label{ah8.equn26}
p=\frac{1}{4} q^{\frac{1}{d-2}} \left[ \frac{(d-3)}{q^{\frac{2}{d-2}}} -\frac{2\Lambda}{d-2} q^{\frac{1}{d-2}} -\frac{\tilde{Q}^2 (d-3)}{q^2}\right] respectively.
\end{equation}
\subsubsection{Thermodynamics in Reduced Phase Space:}
In this section we have studied the Thermodynamics of $d-$dimendional charged AdS BHs in reduced phase space and for this we follow the technique \cite{Baldiotti 2016} given below:\\
Differentiating equation (\ref{ah8.equn25}) and using equation (\ref{ah8.equn26}), we obtain
\begin{equation}\label{ah8.equn27}
dM=\frac{B_{d-2}}{4\pi}\left[\left\{p- \frac{(d-3) \tilde{Q}^2}{4 q^{\frac{2d-5}{d-2}}}\right\}dq-\frac{1}{2(d-1)}q^{\frac{d-1}{d-2}}d\Lambda\right].
\end{equation}
If we assume $\Lambda=\Lambda(q)$, i.e., $\Lambda$ is a function of $q$ only, the equation (\ref{ah8.equn27}) may be expressed as:
\begin{equation}\label{ah8.equn28}
dM=\varpi dq=\frac{B_{d-2}}{4\pi}\left[\left\{p- \frac{(d-3) \tilde{Q}^2}{4 q^{\frac{2d-5}{d-2}}}\right\}-\frac{1}{2 (d-1)}q^{\frac{d-1}{d-2}}\frac{\partial\Lambda}{\partial q}\right]dq.
\end{equation}
The above expression is analogous with the tautological form $\alpha=\varpi dq$ and is restricted to the constraint surface given by the equation (\ref{ah8.equn27}), $dM=\left.\alpha \right|_{\Theta=0}$, where  
\begin{equation}\label{ah8.equn29}
\Theta= p- \frac{(d-3) \tilde{Q}^2}{4 q^{\frac{2d-5}{d-2}}}-\left(\frac{d-3}{4}\right) q^{-\frac{1}{d-2}}+
\frac{\Lambda}{2(d-2)}q^{\frac{1}{d-2}}.
\end{equation}
Here we introduce a symplectic form of $\omega_r$ to represent the expression (\ref{ah8.equn29}) more precisely given by
\begin{equation}\label{ah8.equn30}
\omega_r=dp \wedge \left(\frac{\partial\varpi}{\partial p}\right)dq=dp\wedge \frac{B_{d-2}}{4\pi}dq.
\end{equation}
This equation (\ref{ah8.equn30}) implies that the transformation $(p, q)\mapsto (\varpi, q)$ is canonical.
equation (\ref{ah8.equn27}) may also be written as:
\begin{equation}\label{ah8.equn31}
dM=\Omega^{'} dq-\frac{B_{d-2}}{8\pi (d-1)}\left(q^{\frac{d-1}{d-2}}\Lambda\right)^{'}= \frac{B_{d-2}}{4\pi}\left[\left\{p+ \frac{(2d-5)(d-3) \tilde{Q}^2}{4(d-2) q^{\frac{3d-7}{d-2}}}\right\}-\frac{1}{2 (d-2)}q^{\frac{1}{d-2}}\Lambda\right]dq- \frac{B_{d-2}}{8\pi (d-1)}\left(q^{\frac{d-1}{d-2}}\Lambda\right)^{'},
\end{equation}

where `$^{'}$' stands for the first order derivative. The expression (\ref{ah8.equn31}) shows that there is a time-dependent, i.e., $\tau$-dependent canonical transformation $q\mapsto q, \varpi \mapsto \Omega^{'}$ mapping the reduced phase space with coordinates $(\varpi, q)$ to a reduced phase space with coordinates $(\Omega^{'}, q^{'})$, see detail in reference \cite{Baldiotti 2016}. Hence the tautological form is represented as: 
\begin{equation}\label{ah8.equn32}
d{\cal M}=\Omega^{'} dq^{'}-\frac{B_{d-2}}{8\pi (d-1)}\left(q^{\frac{d-1}{d-2}}\Lambda\right)^{'},
\end{equation} 
where ${\cal M}=M+V\frac{\Lambda}{8\pi}.$ The factor `$-V\frac{\Lambda}{8\pi}$' represents the space time energy.   
For the coordinate transformation in reduced phase space $(\varpi, q)\mapsto (\Omega^{'}, q^{'})$, we consider the generating function of second kind as:
\begin{equation}\label{ah8.equn33}
{\cal F}=\Omega^{'} q^{'}-\frac{B_{d-2}}{8\pi (d-1)}\left(q^{\frac{d-1}{d-2}}\Lambda\right)=\Omega^{'} q+g(q, \tau).
\end{equation}
Differentiating the equation (\ref{ah8.equn33}) with respect to $q$ and substituting $\Omega^{'}$ from equation (\ref{ah8.equn31}), we have 
\begin{equation}\label{ah8.equn34}
\frac{B_{d-2}}{4\pi}\left[\left\{p- \frac{(d-3) \tilde{Q}^2}{4 q^{\frac{2d-5}{d-2}}}\right\}-\frac{1}{2 (d-1)}q^{\frac{d-1}{d-2}}\frac{\partial\Lambda}{\partial q}\right],
\end{equation}
which is equal to $\varpi$, already obtained in equation (\ref{ah8.equn28}). This gives
\begin{equation}\label{ah8.equn35}
\varpi=\frac{\partial \cal F}{\partial q}.
\end{equation}
Moreover from  equation (\ref{ah8.equn33}), we can show
\begin{equation}\label{ah8.equn36}
q^{'}=\frac{\partial \cal F}{\partial \Omega^{'}}=q.
\end{equation}
The canonical transformation (\ref{ah8.equn33}) obey the Bekenstein's prediction on BH entropy and such transformation enforces the second law of BH Thermodynamics. By using the canonical transformation (\ref{ah8.equn33}) one can obtain all the above thermodynamic results directly from Schwarzschild solution without solving the EFE, since in the Schwarzschild mass $\cal M$ the space time energy term which is defined with negative cosmological constant $\Lambda$ is included.

\subsubsection{Thermodynamics in Extended Phase Space:}
One can expand the expression (\ref{ah8.equn30}) in extended phase space with a symplectic form given as:
\begin{equation}\label{ah8.equn37}
\omega_e=dp \wedge \left(\frac{\partial\varpi}{\partial p}\right)dq+d\tau \wedge \left(\frac{\partial\varpi}{\partial \tau}\right)dq+d\zeta\wedge d\tau.
\end{equation}
If $\Lambda$ is taken to be a function of $q$ and $\tau$, i. e., $\Lambda=\Lambda(q, \tau)$, $equation.$ (\ref{ah8.equn28}) can be written as:
\begin{equation}\label{ah8.equn38}
dM=\varpi dq-\frac{B_{d-2}}{8\pi (d-1)}q^{\frac{d-1}{d-2}}\left(\frac{\partial\Lambda}{\partial\tau}\right)d\tau.
\end{equation}
Combining the equations (\ref{ah8.equn12}) and (\ref{ah8.equn14}) with equation (\ref{ah8.equn32}) we have the tautological form $dU\equiv pdq+\zeta d\tau$. Again this form reduces to Poincar\'{e}-Cartan form $dU\equiv pdq-h d\tau$ on the constraint surface $H=0$, where 
\begin{equation}\label{ah8.equn39}
\zeta=-h(q, \tau)=\frac{B_{d-2}}{8\pi (d-1)}q^{\frac{d-1}{d-2}}\left(\frac{\partial\Lambda}{\partial\tau}\right).
\end{equation}
This implies that the symplectic form expressed in $equation.$ (\ref{ah8.equn31}) is true.\\
In coordinates, the Poisson brackets are given as:
\begin{equation}\label{ah8.equn40}
\{f, g\}= \frac{\partial f}{\partial q}\frac{\partial g}{\partial \varpi}-\frac{\partial f}{\partial \varpi}\frac{\partial g}{\partial q}+\frac{\partial f}{\partial \tau}\frac{\partial g}{\partial \zeta}-\frac{\partial f}{\partial \zeta}\frac{\partial g}{\partial \tau},
\end{equation}\label{ah8.equn41}
where $f$ and $g$ are functions on phase space. Hence the non-zero canonical Poisson brackets are
\begin{equation}
\{q, p\}=\{\tau, \zeta\}=1
\end{equation}
Use $equations$ (\ref{ah8.equn15}) and (\ref{ah8.equn24}) to combine with the expression (\ref{ah8.equn28}), we obtain the thermodynamic volume and thermodynamic temperature as:
\begin{equation}\label{ah8.equn42}
V=-\frac{B_{d-2}}{8\pi (d-1)}\left(\frac{4S}{B_{d-2}}\right)^{\frac{d-1}{d-2}}\left(\frac{\partial\Lambda}{\partial P}\right)
\end{equation} and
\begin{equation}\label{ah8.equn42.1}
T_{eff}=T-\frac{B_{d-2}}{8\pi (d-1)}\left(\frac{4S}{B_{d-2}}\right)^{\frac{d-1}{d-2}}\left(\frac{\partial\Lambda}{\partial S}\right).
\end{equation}
Euler's theorem for homogeneous functions $g(x, y)$ such that $g\left(\lambda^{\alpha} x, \lambda^{\beta}y\right)= \lambda^{\gamma} g(x, y)$ turns out to be 
\begin{equation}\label{ah8.equn42.2}
\gamma g(x, y)= \alpha x \left(\frac{\partial g}{\partial x}\right)+ \beta y \left(\frac{\partial g}{\partial y}\right). 
\end{equation}
Taking $g$ as Schwarzschild mass $\cal M$, $x=q$ and $y=0$, Euler's theorem gives
\begin{equation}\label{ah8.equn42.3}
(d-3){\cal M}= (d-2) \Omega^{'} q^{'}.
\end{equation}
In extended space this relation may be expressed as:
\begin{equation}\label{ah8.equn42.4}
(d-3){\cal M}= (d-2) \left(T_{eff}S+ cVP\right),
\end{equation} 
for details see the reference \cite{Baldiotti 2017}.

\section{Maxwell's Equal Area Law And its Extension:}

For Van der Waals equation of state there are also unstable state with $\frac{\partial P}{\partial v}>0$ as charged topological BHs. But they are resolved by using Maxwell's equal area law \cite{Huai 2018} given as: 
\begin{equation}\label{ah8.equn43}
\int_{v_1}^{v_2}P dv.
\end{equation}

Now, we extend this equal area law for $d-$dimensional Charged AdS (Anti-de Sitter) BHs to compute a phase transition of the BH, considered as thermodynamics system.
Using equations (\ref{ah8.equn23}) and (\ref{ah8.equn43}) one can have 
\begin{equation}\label{ah8.equn44}
T ln \left( \frac{v_2}{v_1} \right)- C \left(\frac{1}{v_1}- \frac{1}{v_2}\right)+ \frac{D \tilde{Q}^2}{2d-3} \left( \frac{1}{v_1^{2d-3}}-\frac{1}{v_2^{2d-3}}\right)=P(v_2)-P(v_1).
\end{equation}
Assuming
\begin{equation}\label{ah8.equn45}
P(v_1)= \frac{T}{v_1}-\frac{C}{v_1^2}+ \frac{D \tilde{Q}^2}{v_1^{2(d-2)}}~~~~~
and~~~~~
P(v_2)=\frac{T}{v_2}-\frac{C}{v_2^2}+ \frac{D \tilde{Q}^2}{v_2^{2(d-2)}}
\end{equation}
one can get the right hand side of equation (\ref{ah8.equn44}) as:
\begin{equation}\label{ah8.equn46}
P(v_2)-P(v_1)= T\left(\frac{1}{v_2}-\frac{1}{v_1}\right)-C \left( \frac{1}{v_2^2}-\frac{1}{v_1^2}\right)+ D \tilde{Q}^2 \left( \frac{1}{v_2^{2(d-2)}}-\frac{1}{v_1^{2(d-2)}}\right)
\end{equation}
Equating the relations (\ref{ah8.equn44}) and (\ref{ah8.equn46}) we can establish
\begin{equation}\label{ah8.equn47}
C (xv_2)^{2(d-1)} \left[2(1-x)+(1+x) ln{x}\right]= D \tilde{Q}^2 (xv_2)^{2(d-2)} \left[\frac{2(d-2) \left(1-x^{2d-5}\right)}{2d-5}+ \frac{\left(1-x^{2(d-2)}\right)ln x}{1-x}\right]~,
\end{equation}
where we take $x=\frac{v_1}{v_2} (0<x<1)$.
Again for $P(v_1)=P(v_2)$, one can calculate from equation (\ref{ah8.equn46})
\begin{equation}\label{ah8.equn48}
T (xv_2)^{2d-1}= C (xv_2)^{2(d-1)}(1+x)- D \tilde{Q}^2 (xv_2)^{2(d-2)}\frac{\left(1-x^{2(d-2)}\right)}{1-x}
\end{equation}
If we relate the temperature $T$ with the critical temperature $T_c$ as $T= \xi T_c$, the $equation$ (\ref{ah8.equn48}) takes the form as:
\begin{equation}\label{ah8.equn49}
\xi T_c (xv_2)^{2d-1}= C (xv_2)^{2(d-1)}(1+x)- D \tilde{Q}^2 (xv_2)^{2(d-2)}\frac{\left(1-x^{2(d-2)}\right)}{1-x}~,
\end{equation}
where $T_c = \frac{4 (d-3)^2}{\pi V_c (d-2)(2d-5)}$ with the critical volume $V_c= \frac{4}{d-2} \left[\tilde{Q}^2(d-2)(2d-5)\right]^{\frac{1}{2(d-3)}}$ \cite{Jie 2018}. It is clear that $x=1$ and $\xi= 1$ correspond the critical state of the charged AdS BHs.
\subsection{Approximation Solution:}
We now compute the zeroth order approximation solution at first then substituting this we obtain the first order approximation solution and so on by applying the same method.

The zeroth order approximation solution is calculated as:
\begin{equation}\label{ah8.equn50}
 v_{2;0}^2= F(x)^2=\frac{ D \tilde{Q}^2 \left[\frac{2(d-2) \left(1-x^{2d-5}\right)}{2d-5}+ \frac{\left(1-x^{2(d-2)}\right)ln x}{1-x}\right]}{C x^2 \left[2(1-x)+(1+x) ln{x}\right]}
\end{equation}
and that of first order approximation solution as:
\begin{equation}\label{ah8.equn51}
 v_{2;1}^{2d}= F(x)^{2d}=\frac{ D \tilde{Q}^2 F_(x)^{2d-2}\left[\frac{2(d-2) \left(1-x^{2d-5}\right)}{2d-5}+ \frac{\left(1-x^{2(d-2)}\right)ln x}{1-x}\right]}{C x^2 \left[2(1-x)+(1+x) ln{x}\right]}.
\end{equation}

For $x\rightarrow 1$, $v_2= V_c$, the critical volume of that BH, then from $equations$ (\ref{ah8.equn50}) and (\ref{ah8.equn51}) we can have
\begin{equation}\label{ah8.equn52}
 V_{c}^{2d}=  V_c^{2(d-1)} lim_{x\rightarrow 1} \frac{ D \tilde{Q}^2\left[\frac{2(d-2) \left(1-x^{2d-5}\right)}{2d-5}+ \frac{\left(1-x^{2(d-2)}\right)ln x}{1-x}\right]}{C x^2 \left[2(1-x)+(1+x) ln{x}\right]}
\end{equation}

Meanwhile, from equation (\ref{ah8.equn23}) we can compute
\begin{equation}\label{ah8.equn53}
P= \frac{T}{F(x)}-\frac{C}{F(x)^2}+ \frac{D \tilde{Q}^2}{F(x)^{2(d-2)}}~,
\end{equation}
\section{Clapeyron Equation and The Phase Change:}
Clapeyron equation is the direct experimental verification of some phase changes and is expressed as:
\begin{equation}\label{ah8.equn54}
\frac{dP}{dT}= \frac{L}{T \left(v_{\beta}- v_{\alpha}\right)}~,
\end{equation}
where $L= T \left(s_{\beta}- s_{\alpha}\right)$ is the latent heat of phase change form phase $\alpha$ to phase $\beta$ in which the molar volume and molar entropy are $v_{\alpha}$, $s_{\alpha}$ and $v_{\beta}$, $s_{\beta}$ respectively.

Here we examine the two phase equilibrium coexistence $P$-$T$ curves and the slope $\frac{dP}{dT}$ of that curves for the charged AdS BHs. We may write the equation (\ref{ah8.equn48}) in terms o $F(x)$ as:

\begin{equation}\label{ah8.equn55}
T= G(x)= C\left\{\frac{1+x}{x F(x)}\right\}-D\tilde{Q}^2 \left\{\frac{1-x^{2(d-2)}}{x^3 F(x)^3 (1-x)}\right\}
\end{equation}
And equation (\ref{ah8.equn53}) in terms of $G(x)$ as:
\begin{equation}\label{ah8.equn56}
P= H(x)= \frac{G(x)}{F(x)}-\frac{C}{F(x)^2}+ \frac{D \tilde{Q}^2}{F(x)^{2(d-2)}}.
\end{equation}
Hence one can calculate the slope of two phase equilibrium coexistence $P$-$T$ curves from the equations (\ref{ah8.equn36}) and (\ref{ah8.equn56}) as:
\begin{equation}\label{ah8.equn57}
\frac{dP}{dT}= \frac{H(x)^{'}}{G(x)^{'}}~,
\end{equation} 
where $H(x)^{'}= \frac{dH}{dx}$ and $G(x)^{'}= \frac{dG}{dx}$.
We can have the latent heat of phase change as a function of $x$ for $d-$dimensional charged AdS BH by using the the equations (\ref{ah8.equn54}), (\ref{ah8.equn55}), (\ref{ah8.equn56}) and (\ref{ah8.equn57}) as:
\begin{equation}\label{ah8.equn58}
L=(1-x)\frac{H(x)^{'}}{G(x)^{'}} G(x) F(x).
\end{equation} 
The rate of change of latent heat of phase transition with temperature for $d+1$ dimensional charged AdS BH is obtained from $equations$ (\ref{ah8.equn54}), (\ref{ah8.equn55}) and (\ref{ah8.equn58}) as:
\begin{equation}\label{ah8.equn59}
\frac{dL}{dT}= \frac{dL}{dx} \frac{dx}{dT}=\frac{1}{G(x)^{'}} \frac{dL}{dx}~,
\end{equation}
But for some thermodynamic systems, the rate of change of latent heat of phase transition with temperature is given as \cite{Huai 2018}
\begin{equation}\label{ah8.equn60}
\frac{dL}{dT}= C_{P\beta}-C_{P\alpha}+\frac{L}{T}-\left[\left(\frac{\partial v_{\beta}}{\partial T}\right)_P-\left(\frac{\partial v_{\alpha}}{\partial T}\right)_P \right]\frac{L}{v_{\beta}-v_{\alpha}}
\end{equation}
\begin{figure}[h!]
\begin{center}
~~~~~~~~~~~~~~~~~~Fig.-3~~~~~~~~~~~\\
\includegraphics[scale=.7]{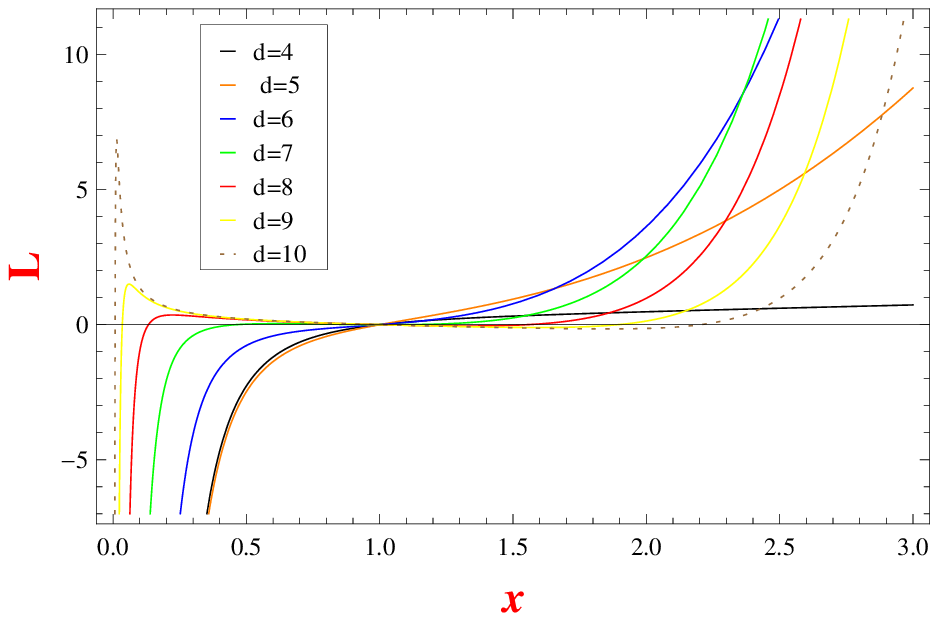}\\

Fig.-3 represents the variations of $ L $ with respect to $x $ for different dimensions $d$.
\end{center} 
\end{figure}

\section{Brief Discussion and Conclusion}
In this letter, we have investigated the thermodynamics of $d-$dimensional ($d>3$) AntideSitter black holes. Here we have considered the thermodynamic aspects of space-times. We have reconstructed the thermodynamic result by using the Hamiltonian approach developed in the reference \cite{Baldiotti 2016}. Initially we have studied the thermodynamics in reduced phase space and correlate with the Schwarzschild solution and then we have extended in the extended phase space. Due to this thermodynamic treatment, the thermodynamic equations of state can be realized as constraints on phase space. In the thermodynamic description of the charged AntideSitter black holes the cosmological constant $\Lambda$ can be introduced as a result of canonical transformation of the Schwarzschild problem. If we considered $\Lambda$ to be the function of coordinates in phase space through, a new thermodynamic equation of state occurs and this new equation of state exhibits the generalized thermodynamic volume. Due to this new approach a new equation of state which completely characterizes the thermodynamics of charged AntideSitter black holes has been established only assuming homogeneity of the thermodynamic variables. We have assumed that $\Lambda$ is a constant in the space-time manifold, though here we considered $\Lambda$ to be the function of coordinates in phase space. It is to be noted that if we vary $\Lambda$, the thermodynamic results which have been developed here remains consistent. In that case $\Lambda$ consider as a dynamical geometric variable \cite{Kastor 2009} rather than the thermodynamic variable.  

When we have plotted the variations of $ P $ with respect to specific volume $\it {v} $ for different dimensions $d$ in Fig.-1a-g for equation (\ref{ah8.equn23}) for different dimensions $d$ for $T<T_c, T=T_c$ and $T>T_c$, we find that there is a small-large black hole phase transition in the system and one can also notice that there are thermodynamic unstable regimes for $\frac{\partial P}{\partial v}> 0$ for $T<T_c$. For higher dimensions, the thermodynamic unstable regimes also higher, i.e., for $T<T_c, T=T_c$ and for some extent of $T>T_c$. The thermodynamic unstable regimes also shown in $P-T$ diagrams in Fig.-2. The negative temperatures for $T<T_c$  as shown in Fig.-2 signifies it. Using ``Maxwell's equal area law", we analyze  the phase transition of the said black holes and the problems mentioned above may be removed. From Fig.-2, we derive the slope of the curves which serve the information about latent heat of phase change by Clapeyron equation which helps to find some thermodynamic properties of thermodynamic system such as black holes and also provide the theoretical background for experimental research on analogous black holes.

 In this letter the main motivations were to establish a thermodynamoc study of the $d$-dimensional black holes. Use of Hamiltinian approach helped us to constract the thermodynamic extended and reduced phases. We observe that for black hole thermodynamics also, these phase spaces have the same results as the classical thermodynamic cosiderations. Two phase equilibrium coexistence is examined and the corresponding latent heat formula if evaluated in Clapeyron equation. As we increase the dimensions, latent heat has more prominent local maxima or minima which indicates the nature of stabilities around the phase transition points. As we increase the position coordinate $x$, i.e., the entropy $S$ to shift from one phase to another more latent heat is to be given if $x$ is low. For medium level of $x$ quantity of latent heat required reduces. But if $x$ is high, monotonically increasing latent heat is required to give. So it is different to change the phase for smaller or larger black holes. The phases are stable then. But for an intermiate site black hole, it will be quite easier to transmite from one phase to another. Some time even we need not to incorporate latent heat from outside. Rather it will itself radiate energy (due to negative latent heat) outward and transmit to another phase. This signifies highly unstable phase. Here the black hole may turn into a hot AdS space. For higher dimension, this unstable region or range of phase space is more prominent. This, on a ather way, depicts that as we increase dimension the one sidal membrane of event horizon becomes more permeable (i.e., unstable) and we are likely to have an unwrapped naked singularity rather than a black hole. This result supports the previously stated outcome of the higher dimensional gravitational collapse \cite{Rudraa 2011, Debnath 2012, Rudrab 2012, Rudrac 2014}, where it was found that if a catastrophic collapse takes place in the background of higher dimensions, naked singularity is likely to occur.

\vspace{.1 in}
{\bf Acknowledgment:}
This research is supported by the project grant of Goverment of West Bengal, Department of Higher Education, Science and Technology and Biotechnology (File no:- $ST/P/S\&T/16G-19/2017$). AH wishes to thank the Department of Mathematics, the University of Burdwan for the research facilities provided during the work. RB thanks IUCAA, PUNE for Visiting Associateship.\\
RB dedicates this article to his PhD supervisor Prof. Subeoy Chakraborty, Department of Mathematics, Jadavpur University, Kolkata-700032, India to tribute him on his $60^{th}$ birth day .

\end{document}